# OPTICAL STOCHASTIC COOLING METHOD IN APPLICATION TO BEAMS OF CHARGED PARTCLES

E.G. Bessonov (FIAN, Moscow, Russian Federation),
A.A. Mikhailichenko (Cornell University, CLASSE, Ithaca, NY 14853, U.S.A.)

**ANNOTATION**

We discuss the Optical Stochastic Cooling (OSC) method in applications to the beams of charged particles, circulating in accelerators and storage rings. In this publication we concentrated on various OSC schemes in a diluted beam approximation, when the heating of selected particle by its neighboring ones could be neglected. Even so, this approximation allows us to identify important features in the beam cooling. In the forthcoming publication, on the basis of approach developed here, we will include effects of heating in the dynamics of cooling.

## 1. INTRODUCTION

The term "particle beam cooling" means reduction in the six-dimensional phase space (emittance) occupied by this beam. The smaller the beam emittance is and the greater the number of particles in it, the less the product of the geometrical sizes and quantities of momentum spread in the beam– brighter is the light source based on this beam, and higher the luminosity of the colliding beams is in experiments on high-energy physics. This is a relevance of the topic of cooling.

The most general scheme of optical stochastic cooling of charged particles beams is shown in Fig. 1. It uses a signal undulator (pickup), an optical line with adjustable delay, an optical amplifier (OA), the optical filters, lenses, a mobile screen, one or more corrective undulators (kickers). In the area between the pickup and kicker undulators (bypass) the magnetic structure used which allows a beam passage with the lowest possible distortion of its spatial structure. In the optical pickup undulator of cooling system the relativistic particles emit electromagnetic waves in the form of undulator wave packets (UWP) with the number of periods equal to the number of undulator periods, $M$. These packages are moving in the direction of undulator axis, then enhanced by the broadband OA and enter the kicker undulator simultaneously with the particles emitted their UWP. In this kicker undulator the UWP amplified is correcting the particle trajectory.

Fast movable screen, which is located in the image plane of the beam, produces selection (screening) of unnecessary parts of UWP, which now becomes a function of particle's energy and deviation from its instantaneous radial orbit. The delay line is used for arrangement of congruence of the beam of particles with UWP at required phase (phase delay).

In this paper, a unified description covers the basics of theory of optical stochastic cooling methods of particle beams (electrons, ions, muons) in storage rings and discusses theirs advantages and disadvantages. Substantially in the optical stochastic cooling techniques the idea of a conventional microwave stochastic cooling further developed [1]. Practically in all publications on optical stochastic cooling the undulators is proposed to use as a signal and control devices (pickup and kicker)[1]. The method of optical stochastic cooling (OSC) has been proposed in [2]. Method of transit time options of OSC optical cooling (TTOSC) was developed in [3]. In [4] a method was proposed for enhanced (rapid) optical cooling (EOC). In the method of OSC it was proposed to use a quadrupole undulator as a signal pickup, and in TTOSC and EOC methods - conventional (dipole) undulators are in use. In the EOC method it was proposed to use a special system for selection of UWP by a movable screen which makes parameters of UWP sensitive to the sign of particle's deviation from its instantaneous orbit, while entering the pickup undulator. Here it was drawn attention to the importance of taking into account the quantum nature of the emission of UWP and to the grouping of particles in equilibrium regions with different energies [4].

---

[1] In publication [19] the gratings are used instead.

The main feature of the methods of optical stochastic cooling in comparison with the classical scheme [1], is that in the optical methods the signal (UWP) does not bear a DC component and it is a harmonically varying function of time (longitudinal coordinate) at distances much smaller, than the bunch length. By analogy with the classical scheme [1], the particle beam cooling occurs due to the interaction of particles with theirs own amplified UWP in the field of kicker undulator, leading to a change in the energy of these particles (see item 3). For doing this, the kicker undulator is located at the place of orbit, where there is a dependence of the transverse position of particle on the energy (dispersion), see Figure 2.

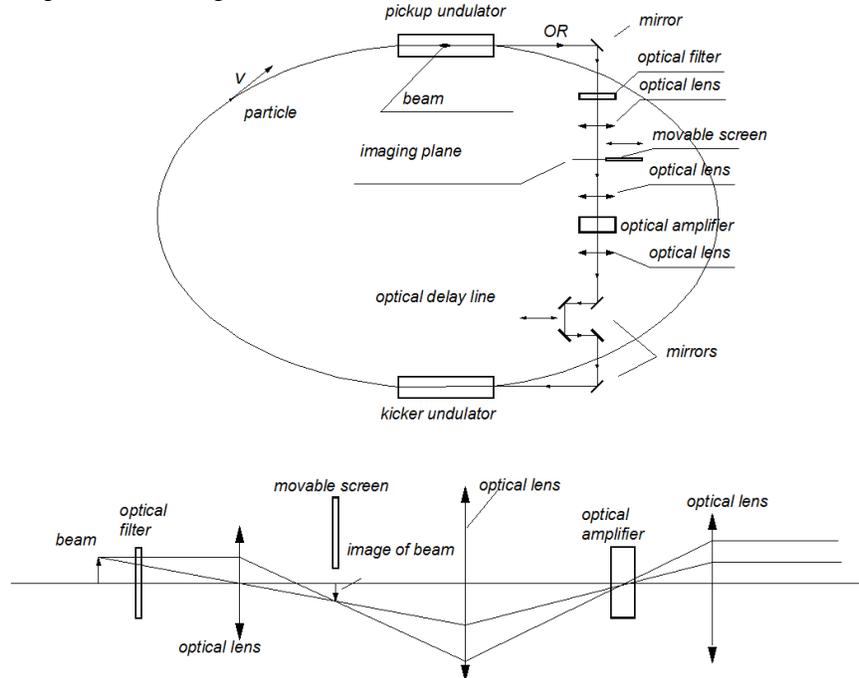

**Figure 1.** Scheme of optical cooling (above) and its disclosed optical structure (below).

It is thus seen that the circuit of (optical) stochastic cooling is some kind of feedback serving the loop: beam- amplifier-beam.

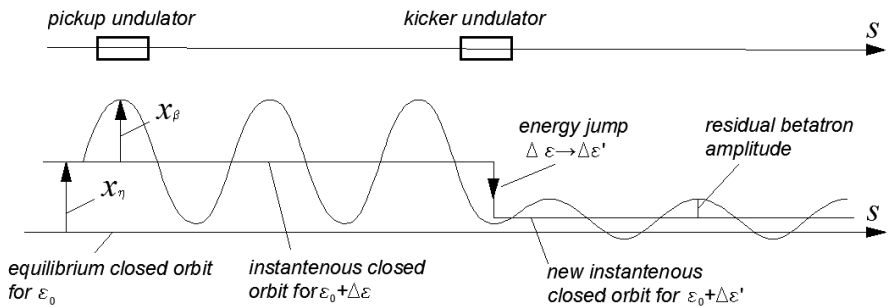

**Figure 2.** The way to arrange a transverse kick by changing the energy of particle. Such phasing of betatron oscillations and dispersion function delivers simultaneous cooling of betatron oscillations and the energy spread.

Let us make few preliminary remarks. In any method of stochastic cooling, including an optical one, the bandwidth in which the amplifier operates plays a key role [1] - [5]. If the microwave amplifier in microwave stochastic cooling system overlaps bandwidth of several GHz, the optical bandwidth can be expected $10^4$ times greater. This leads to the fact that it is possible to distinguish the structure of the bunch on the distances of a few micrometers, thereby minimizing the influence of neighboring particles on any selected one and to accelerate the cooling process significantly. Thus the average number of particles involved in the overlapping of their UWPs is $N_S \sim N_B c / (\Delta f \cdot l_B) = N_B l_{UWP} / l_B$, where $N_B$ – is the total number of particles, $\Delta f$ - is the bandwidth of UWP and OA, $l_{UWP}$ - is UWP length, $l_B$ - is the bunch length[2], $c$-is the speed of light.

In this publication we consider the formalism OSC in approximation $N_S \sim 1$. Formally, for $N_S \sim 1$ the number of turns required for damping of transverse oscillations is $\sim 1$, but the amplification of UWP in optical amplifier in this case becomes non practical, so it requires a many-turn damping anyway. Even more–despite of this simplification, it is possible to identify the set of very interesting details of the cooling process and to understand the peculiarities of different variants of OSC. These are: self-stimulated radiation in undulator, described first in [6], [7], quantum effects reflecting the peculiarities of radiation when the amount of energy radiated in undulator is not enough to build up the energy of quanta [4], [8]. Also, there was attracted attention to the fact, that the energy of quanta radiated by a particle in undulator is $\sim E_\perp^2$, while the kick is $\sim E_\perp$, where $E_\perp$ - is the strength of electromagnetic wave in UWP.

Complete presentation of the cooling process, corresponding $N_S > 1 \div 10$, will be presented in a forthcoming publication.

## 2. BASICS FROM CYCLIC ACCELERATORS

In circular accelerators and storage rings the charged particles move in closed annular magnetic systems consisting of bending magnets separated by the straight sections, which contains the focusing lenses, the injection and ejection systems, radio frequency (RF) accelerating cavities, beam observation system, undulators and other equipment. The vacuum chamber running through all the elements of the magnetic system separates the region occupied by the beam from the outer space.

The motion of particles is described in a curvilinear coordinates *x, z, s*, where *x, z* – are the transverse and *s* – is the longitudinal coordinate measured along the selected instant (closed) orbit from some start point. Typically the basis of selected ideal instantaneous plane orbit is located in a horizontal plane. Real distortions of the orbit calculated on the base of measurements of magnetic fields generated by the bending and focusing elements of structure as well as on drive through numerous surveillance methods of circulating beam. In the bending magnets the particle trajectories are close to a circle, and in linear sections - to the straight lines. Circumference of instant particle orbit *C* is proportional to their current momentum *p*.

The equilibrium particle orbit is a such instantaneous orbit at which the frequency of revolution $\omega_s = C_s / v_s$ is equal or less than an integer multiple of accelerating RF frequency, $\omega_{RF}$. Hereinafter, an index *s* value designated to equilibrium, and *v* -is the speed of particles.

Motion of particle in a storage ring is described by equations having periodic coefficients. The period may be the associated with the instantaneous circumference of the orbit of the particles *C*. If the magnetic structure has a symmetry of the order *m*, then the length of period is equal to *C/m*. In the horizontal plane the trajectories of particles can be described as sum of two functions, one of

---

[2] For continuous beam the perimeter of orbit should be substituted instead of beam length.

which- $x_\eta(s)$ describes the deviation of the instantaneous orbit from the equilibrium one at the position of the particle, and the other $x_\beta(s)$ - the betatron oscillations of particle relative to the instantaneous orbit:

$$x(s) = x_\beta(s) + x_\eta(s), \qquad (1)$$

where $s = vt$ - is the path traversed by the particle along the instantaneous orbit during time $t$. In the vertical direction the deviation of particle with respect to the instantaneous orbit denotes by function $z_\beta(s)$. The values $x_\beta(s)$ $z_\beta(s)$ are described by the Hill equation and expressed in terms of periodic $\beta$ – functions $\beta_{x,z}(s) > 0$, and $x_\eta(s)$ - through a dispersion function $\eta_x(s)$ in the form

$$x_\beta(s), z_\beta(s) = \sqrt{\beta_{x,z}(s)}[a_{x,z}\cos\mu_{x,z}(s) + b_{x,z}\sin\mu_{x,z}(s)], \quad x_\eta(s) = \eta_x(s)\delta, \qquad (2)$$

where $a_{x,z}$, $b_{x,z}$ -constants defined by the initial conditions, $\mu_{x,z}(s) = \int_0^s ds'/\beta_{x,z}(s')$, $\delta = \Delta p/p \cong \beta^{-2}\Delta\varepsilon/\varepsilon$, $\beta = v/c$, $c$ - is a speed of light, $p$ - momentum, $\Delta p$ - deviation of momentum from its equilibrium value $p_s$, $\Delta\varepsilon$ - deviation of particle's energy from the equilibrium one $\varepsilon_s$. For any storage ring the functions $\beta_{x,z}(s)$ and the functions

$$\eta_x(s) = \frac{\sqrt{\beta_x(s)}}{2\sin(\pi v_x)}\int_s^{s+C}\frac{\sqrt{\beta_x(s')}}{R(s')}\cos[\mu_x(s') - \mu_x(s) - \pi v_x]ds', \qquad (3)$$

could be calculated analytically (in simplest case) or numerically and could be represented as tables and graphs [9] - [11]. Here $v_{x,z} = (1/2\pi)\int_0^C ds/\beta_{x,z}(s) = \mu_{x,z}(C)/2\pi$ - is the number of betatron oscillations per revolution committed by particle, $R(s)$ is the instant bending radius in the magnets. The deviations of particle $x_\beta(s)$ and $z_\beta(s)$ in (2) can be conveniently represented in the form

$$x_\beta(s), z_\beta(s) = A_{x,z}(s)\cos\phi_{x,z}, \qquad (4)$$

where $A_{x,z}(s) = \sqrt{(a_{x,z}^2 + b_{x,z}^2)\beta_{x,z}(s)}$ and $\phi_{x,z} = \alpha_{x,z} - \mu_{x,z}(s)$ - are the local amplitude and phase of the particle betatron oscillations, $\alpha_{x,z} = \arccos(a_{x,z}/\sqrt{a_{x,z}^2 + b_{x,z}^2})$. Magnitudes $A_{x,z}(s)$ and $\alpha_{x,z}$ - determined by the initial conditions. Phase $\phi_{x,z}$ is conveniently written as $\phi_{x,z} = 2\pi s/\lambda_{x,z} + \chi_{x,z}(s)$ where $\lambda_{x,z} = C/v_{x,z}$ - wavelength of betatron oscillations, $\chi_{x,z}(s) = \alpha_{x,z} - \mu_{x,z}(s) - 2\pi s/\lambda_{x,z}$, $\chi_{x,z}(s) = \chi_{x,z}(s + C/m)$.

The origin of signal arranged at the entrance of the undulator. In this case, the deviation of particle at this place on the *n*-th turn, according to (4) has the form

$$x_\beta(n,s) = A_x(0)\cos[\alpha_x - 2\pi n v_x - \mu_x(s)], z_\beta(n,s) = A_z(0)\cos[\alpha_z - 2\pi n v_z - \mu_z(s)] \; (n=0,1,2,). \; (5)$$

There are two modes of cooling beams of particles in the storage rings. These modes are switched *off/on* the accelerating RF voltage across the gap of the accelerating cavity drive. *Off* mode RF voltage can be used for cooling of heavy particles (protons, muons, charged ions) when synchrotron radiation of particles could be neglected. Particles in the *on* mode accelerating RF voltage driving the resonator make stable phase oscillations in the longitudinal phase space bounded by a separatrix. They are grouped around the synchronous particle moving with the frequency of RF voltage applied across the gap of resonator. RF cavity is out of the cooling system of the beam. Possible is usage of two cooling circuits. One of these is concentrated on rapid cooling rate in the longitudinal and the second one- in the transverse directions.

## 3. BASICS FROM UNDULATOR RADIATION

Undulator -is a device which is forming a periodic static or a time-varying electromagnetic field along an axis. Charged particles are moving in a direction close to the axis of the undulator, perform in this field periodic vibrational reciprocating motion, i.e. represent the oscillators which are moving with an average velocity $|\vec{\overline{v}}| < v$, $\vec{\beta} = \vec{\overline{v}}/c$ where $\vec{\overline{v}}$ - the vector of mean particle velocity. Such oscillators in accordance with the relativistic Doppler effect, emit an undulator radiation in a direction of their average velocity, with the carrier frequency coinciding with the first harmonic, which is $\omega_{1c} = 2\pi c / \lambda_{1c}$ in the relativistic case and the corresponding wavelength

$$\lambda_{1c} \cong \frac{\lambda_u}{2\gamma^2}(1 + \overline{p_\perp^2}) \qquad (6)$$

where $\lambda_u$ - is period of undulator, $\gamma$ - is a relativistic factor, $\overline{p_\perp^2} = \gamma^2 \overline{\beta_\perp^2}$ - is a dispersion of relative transverse momentum, $\overline{\beta_\perp^2}$ is - the dispersion of the transverse relative velocity of the particle, $\beta_\perp = |\vec{\beta}_\perp|$, $\vec{\beta}_\perp = \vec{v}_\perp / c$, $\vec{v}_\perp$ is the vector of particle's transverse velocity [12] - [14]. Modern literature $\overline{p_\perp^2}$ is often denoted by the value $K^2$, and the value of a parameter $K = \sqrt{\overline{p_\perp^2}}$ called the power of the undulator.

The trajectories of particles in undulators typically represent the form of sine waves or spirals. Transverse oscillations of the particles in the aggregate field of kicker undulator and in the field of UWP, amplified in the OA, changes the energy of the particle. Thus, in a simplest case of helical trajectories in the helical undulator and a circularly-polarized wave, the energy change becomes

$$\Delta E = e\int_0^T (\vec{v}\vec{E})\,dt = e\int_0^T (\vec{v}_\perp \vec{E}_\perp)\,dt \cong eKL_u E_\perp Cos\varphi / 2\gamma$$

where - $\Omega = 2\pi c / \lambda_u$, $L_u = M\lambda_u$ -is the length of kicker undulator, $\varphi$ -is the entry phase of the particle in the UWP.

## 4. BASICS FROM THE TRANSIT-TIME METHOD IN OSC (TTOSC)

Let us call the reference (refer) particle, which has a predetermined energy $\varepsilon_r$, zero amplitude of betatron oscillations and which enters the kicker undulator simultaneously with its emitted UWP. Such a particle moves along a closed orbit of the reference, has zero phase and, in the absence of screening (case OSC) - the maximum rate of energy loss by [4]

$$P_{loss}^{max} = \frac{8\pi\sqrt{\pi}e^2 f_r \Phi(N_{ph}^{cl}) N_{kick} K^2 \sqrt{\alpha_{ampl}}}{(1+K^2)\lambda_{1,min}}, \qquad (7)$$

where $\alpha_{ampl}$ - is an amplification of OA, function $\Phi(N_{ph}^{cl})\big|_{N_{ph}^{cl}\sim 1} = \sqrt{N_{ph}^{cl}}\big|_{K=1,Z=1} \sim 0.107$, $N_{ph}^{cl} = \pi\alpha Z^2 K^2 / (1+K^2)$ - is the number of photons with energy $\sim \hbar\omega_{1c}$ in UWP in the bandwidth $\Delta\omega/\omega \approx 1/M$, $\Phi(N_{ph}^{cl} > 1) = 1$, $f_r$ -is a revolution frequency of particle, $N_{kick}$ - is the

number of kicker undulators, [4], [12]-[14]. In general, the imaginary particle can serve as a reference one if there is no particle with required energy and direction of its velocity.

### A. Beam cooling with RF turned off.

If the RF drive system is off, the radiative energy loss of the particles can be neglected, no screening, the rate of change of the particle energy in the drive during its interaction with its own UWP in a kicker undulator is determined by the average power

$$\bar{P}_{loss} = -P_{loss}^{\max} f(\varphi_{in}) \langle \cos \varphi_{in} \rangle, \tag{8}$$

where $\varphi_{in} = \varphi_{in,\eta} + \varphi_{in,\beta}$ - is the inlet phase of particle with respect to UWP at the entrance in the kicker undulator. The magnitude $\varphi_{in,\eta} = \omega_{1c} \Delta t_\eta$ is the phase shift of the particle at the kicker entry; this particle does not commit betatron oscillations due to the deviation of its energy from the reference energy $\Delta \varepsilon_{in} = \varepsilon_{in} - \varepsilon_r$, $\Delta t_\eta$ -is the difference of times at the entrance of the kicker undulator between the transit time of the particle itself and the reference path time of its amplified UWP. The phase $\varphi_{in,\beta} = \omega_{1c} \Delta t_\beta$ is due to the betatron oscillations of particles, $\Delta t_\beta$ -is the difference between the times of passing particles, with and without betatron oscillations on the ways from the entrance into the signal (pickup) undulator down to the entering into the kicker undulator $f(\varphi_{in})|_{|\varphi_{in}|<2\pi M} = 1 - |\varphi_{in}|/2\pi M$, $f(\varphi_{in})|_{|\varphi_{in}|>2\pi M} = 0$. Sign <> denotes averaging in (7) over large number $n$ of particle passages through the pickup undulator; it takes into account dependence of the magnitude and sign of the phase jumps of the particle phase $\varphi_{in,\beta}(n)$ at the entry in UWP due to betatron oscillations of the particle at the entrance to the undulator signal (see below).

Energy deviation from the reference particle's energy and the betatron oscillations of particle alter the length of it's path between the signal and kicker undulators $L_{p,k,0}$ on the value $\Delta L_{p,k} = \Delta L_{p,k,\eta} + \Delta L_{p,k,\beta}$ and the corresponding shift of the initial phase at the entrance into the kicker undulator $\Delta \varphi_{in} = \Delta \varphi_{in,\eta} + \Delta \varphi_{in,\beta}$ regarding the reference phase particles. Here $\Delta \varphi_{in,\eta} = \omega_{1c} \Delta t_\eta = k_{1c} \Delta L_{p,k,\eta}$ -is the phase shift due to the deviation of energy (momentum) of a particle with respect to the reference energy, $\Delta \varphi_{in,\beta} = \omega_{1c} \Delta t_{\beta,x} = k_{1c} \Delta L_{p,k,\beta}$ - phase shift caused by its amplitude of betatron oscillations, $k_{1c} = \omega_{1c}/c$.

The increment of the path length $L_{p,k}$ between the signal and kicker undulator, due to the deviation from the reference energy of the particle energy and betatron oscillations of the particles is given by

$$L_{p,k} = \int_0^{L_{p,k,0}} \sqrt{x_\beta'^2 + z_\beta'^2 + [1 + (x_\beta + x_\eta)/\rho]^2} \, ds \cong$$
$$\int_0^{L_{p,k,0}} \{1 + 0.5[x_\beta'^2 + z_\beta'^2 + 2(x_\beta/\rho) + 2(x_\eta/\rho) + (x_\beta/\rho)^2 + (x_\eta/\rho)^2]\} ds, \tag{9}$$

$$\Delta L_{p,k} \cong \tfrac{1}{2} \int_0^{L_{p,k,0}} [x_\beta'^2 + z_\beta'^2 + 2x_\beta/\rho + (x_\beta/\rho)^2 + 2x_\eta/\rho + (x_\eta/\rho)^2] ds, \tag{10}$$

where $x_\beta' = dx_\beta/ds$, $z_\beta' = dz_\beta/ds$, $x_\beta$ and $z_\beta$ - are the transverse coordinates of particle, $\rho = \rho(s)$ - is the local radius of curvature in bending magnets [9] - [11].

Deviations of sought entry phase of the particles in their emitted and amplified UWP at the entrance into the kicker undulator, according to expressions (4), (9) and $x_\eta(s) = \eta_x(s)\delta$ have the form

$$\Delta\varphi_{in,\eta} = k_{1c}\int_0^{L_{p,k,0}} \{x_\eta(s)/\rho + 0.5[x_\eta(s)/\rho]^2\}ds = \eta_{c,l}^1\delta + \eta_{c,l}^2\delta^2,$$

$$\Delta\varphi_{in,\beta} = k_{1c}\int_0^{L_{p,k,0}} \{x_\beta(s)/\rho + 0.5[x_\beta'^2 + z_\beta'^2 + (x_\beta/\rho)^2]\}ds, \quad (11)$$

where $\eta_{c,l}^1 = k_{1c}\int_0^{L_{p,k,0}}(\eta_x/\rho)ds$ и $\eta_{c,l}^2 = 0.5k_{1c}\int_0^{L_{p,k,0}}[x_\eta(s)/\rho]^2 ds$ - are local slippage factors of the first and second order respectively. From the expression $\Delta\varphi_{in,\eta} = \omega_{1c}\Delta t_\eta = \eta_{c,l}^1\delta + \eta_{c,l}^2\delta^2$ in the first approximation it follows, that $\omega_{1c}T_{p,k}(\Delta t_\eta/T_{p,k}) = \eta_{c,l}^1(\Delta p/p)$ or, as $\Delta t_\eta = \Delta T_{p,k}$, $\omega_{1c}T_{p,k}\eta_{c,l} = \eta_{c,l}^1$, where $\eta_{c,l} = d\ln T_{p,k}/d\ln p$ - is a similar slippage factor, but with a different coefficient [4].

Phase deviation $\Delta\varphi_{in,\beta}$, according to (5), in the first approximation can be written as

$$\Delta\varphi_{in,\beta}^1 = [k_{1c}/\sqrt{\beta_x(s)}]\int_0^{L_{p,k,0}}[A_x(0)\sqrt{\beta_x(s)}/\rho]\cos[\alpha_x + 2\pi n\nu_x - \mu_x(s)]ds =$$

$$[k_{1c}A_x(0)]\{\cos[\alpha_x + 2\pi n\nu_x]\int_0^{L_{p,k,0}}(1/\rho)\cos[\mu_x(s)]ds + \sin[\alpha_x + 2\pi n\nu_x]\int_0^{L_{p,k,0}}(1/\rho)\sin[\mu_x(s)]ds\} =$$

$$\Delta\varphi_{in,\beta,m}^1 \cos[\alpha_x + b_x + 2\pi n\nu_x], \quad (12)$$

where $\Delta\varphi_{in,\beta,m}^1 = k_{1c}A_x(0)B_x(0)$, $b_x = c_x/\sqrt{c_x^2 + d_x^2} \leq 1$, $B_x(0) = \sqrt{c_x^2 + d_x^2}$,

$$c_x = [1/\sqrt{\beta_{x,z}(0)}] \times \int_0^{L_{p,k,0}}(\sqrt{\beta_{x,z}(s)}/\rho)\cos[\mu_x(s)]ds,$$

$$d_x = [1/\sqrt{\beta_x(0)}]\int_0^{L_{p,k,0}}(\sqrt{\beta_x(s)}/\rho)\sin[\mu_x(s)]ds.$$

From the expressions (11) and (12) it follows that the entry phase deviation of particle $\Delta\varphi_{in}$ depends smoothly on its energy and on the amplitude of betatron oscillations. In this case it is done very differently in the magnitude and sign of jumps from the turn number to turn over number of particle $n$. For example, the introduction in a cooling system an additional bending magnet with the length $L_{BM} \ll \lambda_x$, in accordance with (11), leads to a change in the phase jump, and the energy of particle passing through it when a deviation in amplitude close to a first approximation, becomes equal to

$$\left|\Delta\varphi_{in,\beta}^1\right| \sim k_{1c}A_x(s_{BM})L_{BM}/\rho. \quad (13)$$

$$\Delta\varepsilon_{jp} = P_{loss}T = -P_{loss}^{max}T f(\varphi_{in})\cos(\varphi_{in}). \quad (14)$$

Here $T$ –is the period of revolution in a storage ring, $\varphi_{in}(n) = \varphi_{in,\eta} + \Delta\varphi_{in,\beta}(n)$. For $L_{BM} = \rho/10^2$, $\lambda_{1c} = 0.5\,\mu m$, $A_x(0) = 1\,mm$ the value $\Delta\varphi_{in,\beta} \sim 40\pi \gg 1$. If $|\Delta\varphi_{in,\beta}| \geq \pi/2$ then the value and sign of jumps of energy could change from turn to turn negatively affects the cooling rate of the peripheral particles. Therefore, the portion of the magnetic structure of the cooling system drive, bypass, according to (11), it is desirable to approach the isochronous motion by betatron oscillations ($\Delta\varphi_{in,\beta}^1 < 1$ or $B_x(0) \sim c_x \sim d_x \ll 1$).

In a second smooth approximation ($x_\beta, z_\beta = A_{x,z}\cos[k_{x,z}s - 2\pi n\nu_x]$, $k_{x,z} = 2\pi/\lambda_{x,z}$, $|\rho| = const.$) the phase shift becomes

$$\Delta\varphi_{in,\beta}^2 = 0.5k_{1c}\int_0^{L_{p,k,0}}[x_\beta'^2 + z_\beta'^2 + (x_\beta/\rho)^2]ds =$$

$$0.5\pi k_{1c}k_{x,\beta}A_x^2(0)(1 + 1/k_{x,\beta}^2\rho^2) + 0.5\pi k_{1c}k_{z,\beta}A_z^2 \quad (15)$$

Hence, for example, in a typical case $\lambda_{1c} = 0.5$ microns, $L_{p,k} \sim \lambda_x$, $\lambda_x \sim \lambda_z \sim 10 m$, $k_{x,\beta}\rho = 1$, $A_x = A_z = 1 mm$ the value $\Delta\varphi^2_{in,\beta} \sim 20$. This example shows that the value of (14) also imposes stringent requirements on the value of the amplitude of betatron oscillations. Note that in the second approximation the phase jumps have the same sign and are equal in magnitude, i.e. do not depend on the turn number *n*.

### *a. Beam cooling in a longitudinal phase space.*

The rate of power loss by particles (7) for small phase shift $\varphi_{in,\beta} \ll 1$ is an oscillating function of deviation of the particle energy from the reference one $\varepsilon - \varepsilon_r$ (see Fig. 3). Its relative amplitude $f(\varphi_{in})$ decreases linearly from the maximal $f(\varphi_{in} = 0) = 1$ to a zero value at the corresponding phase $\varphi_{in} = \varphi_{in,\eta} = \omega_{1c}\Delta t = \pm 2\pi M$, which corresponds to the deviation of the particle energy $|\varepsilon - \varepsilon_r| \geq M \cdot \delta\varepsilon_{gap}$, where the size of the energy gap between the energies of the stable equilibrium of particles is

$$\delta\varepsilon_{gap} \sim 2\pi\beta^2\varepsilon_\gamma / \eta^1_{c,l}. \qquad (16)$$

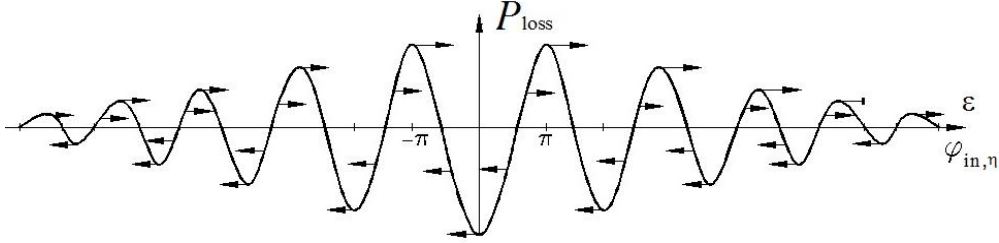

**Figure 3.** In optical cooling the particles are grouping around the phases
$\varphi_{in}(\delta_m) = -\pi/2 + 2\pi m$ (with energy $\varepsilon_m$)

Under these conditions, the particles located in the range of phases around $-\pi/2 + 2m\pi \leq \varphi_{in} \sim \varphi_{in,\eta} \leq \pi/2 + 2m\pi$ monotonically losing theirs energy, and the particles are located within the range of phases $-3\pi/2 + 2m\pi \leq \varphi_{in} \sim \varphi_{in,\eta} \leq -\pi/2 + 2m\pi$ - are accelerated, and as a result grouped near *2M* equilibrium phases [4]

$$\varphi_{in,\eta,m} = -\pi/2 + 2m\pi, \quad m = 0, \pm 1, \ldots \pm (M-1), \quad (\varphi_{in,\beta} = 0). \qquad (17)$$

Some particles can be trapped near the phase of unstable equilibrium $\varphi'_m = \pi/2 + 2m\pi$, since in this area the rate of change of the particle energy $P_{loss}(\varphi_m) \sim 0$. Phases $\varphi_m = \varphi_{in,\eta,m}$ ($\varphi_{in,\beta} \sim 0$, $A_{x,z} \sim 0$) in (17), according to (16) correspond to the equilibrium energy [4]

$$\varepsilon_m = \varepsilon_r[1 + \frac{2\pi m\beta^2}{\eta^1_{c,l}}]. \qquad (18)$$

Thus, if particles in the beam have a small spread of the betatron amplitudes $A_x, A_z$ ($\varphi_{in,\beta} \ll 1$) and large energy spread, they are grouped into 2*M* energy zones whose centers are determined by the equilibrium phases $\varphi_m$ and energies $\varepsilon_m$. The depth of the energy bands decreases with increasing the numbers $|m|$. In each energy zone the cooling of beam is going in the longitudinal phase space. Cooled beam of particles becomes overwhelmed by $2M$ quasi-monochromatic layers of energy.

Phase shift $\Delta\varphi_{in,\beta}$ determined by expression (11) is the sum of the fixed and variable components. Full phase shift in the second approximation in the amplitude of the betatron oscillations can be represented in the form $\Delta\varphi_{in}(n) = \Delta\varphi_{in,\eta} + \Delta\varphi^1_{in,\beta}(n) + \Delta\varphi^2_{in,\beta}$ where the variable component $\Delta\varphi^1_{in,\beta}(n)$ varies both in magnitude and sign. Average $<\Delta\varphi^1_{in,\beta}(n)>=0$.

The rate of change of the particle's energy in the storage ring, according to (7), is proportional to the magnitude $\langle\cos\varphi_{in}\rangle = (1/N)\sum_{n=1}^{N}\cos[\varphi_{in,\eta} + \Delta\varphi^1_{in,\beta}(n) + \Delta\varphi^2_{in,\beta}]$ where the averaging is going over the number of revolutions $N \gg 1$. If the sum divided into pairs, in which the terms have the same magnitude and different sign betatron phase $\Delta\varphi^1_{in,\beta}(k) = -\Delta\varphi^1_{in,\beta}(l) > 0$, the sum of each pair will have the form

$$\cos[\varphi_{in,\eta} + \Delta\varphi^1_{in,\beta}(k) + \Delta\varphi^2_{in,\beta}] + \cos[\varphi_{in,\eta} + \cos[\varphi_{in,\eta} + \Delta\varphi^1_{in,\beta}(l) + \Delta\varphi^2_{in,\beta}] =$$
$$= 2\cos(\varphi_{in,\eta} + \Delta\varphi^2_{in,\beta})\cos\Delta\varphi^1_{in,\beta}(k),$$

and the average value of the sum takes the form

$$\langle\cos\varphi_{in}\rangle = (1/K)\cos(\varphi_{in,\eta} + \Delta\varphi^2_{in,\beta})\sum_{k=1}^{K}\cos[\Delta\varphi^1_{in,\beta,m}\cos(\alpha_x + b_x + 2\pi k\nu_x)], \quad (19)$$

where $K=N/2$ -is the number of pairs of instantaneous jumps of orbit. Averaging procedure in this expression can be simplified, leaving remnants of the arguments of its phases $\alpha_x + b_x + 2\pi k\nu_x = \xi_k + 2\pi m$ which are $\xi_k \leq 2\pi$, placing them in the range of phases $-\pi/2 \leq \xi_k \leq \pi/2$, i.e. where $\Delta\varphi^1_{in,\beta}(k) = \Delta\varphi^1_{in,\beta,m}\cos\xi_k > 0$, at intervals $d\xi = \pi dk/K$ so that the values in ascending order. In this case, the sum of terms in (18) can be replaced by an integral

$(1/\pi)\int_{-\pi/2}^{\pi/2}\cos[\Delta\varphi^1_{in,\beta,m}\cos\xi]d\xi = J_0(\Delta\varphi^1_{in,\beta,m})$, where $J_0(\Delta\varphi^1_{in,\beta,m})$ - is Bessel function of zeroth order.

Thus, the average rate of change of the particle energy and the average energy jumps of particles in the storage ring for the arbitrary values of betatron amplitudes take the form

$$\overline{P}_{loss} = -P_{loss}^{max}f(\varphi_{in})\cos(\varphi_{in,\eta} + \Delta\varphi^2_{in,\beta})J_0(\Delta\varphi^1_{in,\beta,m}), \quad \overline{\Delta\varepsilon}_{jp} = -\overline{P}_{loss}T \quad (20)$$

From here it follows that when $J_0(\Delta\varphi^1_{in,\beta,m}) > 0$, i.e. at amplitudes of particle phase in the UWP at the entrance into the kicker undulator satisfying condition $0 < \varphi^1_{in,\beta,m} < \mu_1$, $\mu_{2+2i} < \varphi^1_{in,\beta,m} < \mu_{3+2i}$, $i = 0,1,2,3...$ in the areas of energy $m$, there is going cooling of the particle beam in longitudinal phase space (energy). Here - $\mu_j$ - is the root of the Bessel function. The particles become collected at phases

$$\varphi_{in,\eta,m} = -\pi/2 - \Delta\varphi^2_{in,\beta} + 2m\pi \quad (\varphi_{in,\beta} = 0). \quad (21)$$

From (21) it follows that in absence of decrement of betatron amplitudes of the particles or in case of absence of increment, the buildup phase $\varphi_{in,\eta}$ does not reach its equilibrium position $\varphi_{in,\eta,m} = -\pi/2 + 2m\pi$ on the amount $\Delta\varphi^2_{in,\beta}$ or, instead, it becomes increased. Therefore, in the limit of the beam spread of the particles remains appropriate energy.

Therefore, in the limit, the appropriate beam energy spread of the particles remains as big as

$$\Delta \varepsilon = \varepsilon_r \Delta \varphi_{in,\eta} / \eta_{c,l}^1 = \varepsilon_r \Delta \varphi_{in,\beta}^2 / \eta_{c,l}^1. \quad (22)$$

For the parts of the transverse phase space corresponding to the amplitude of the betatron phases satisfying $J_0(\Delta \varphi_{in,\beta,m}^1) < 0$ the cooling of beam energy is going in other areas having centers around

$$\varphi_{in,\eta,m} = \pi / 2 - \Delta \varphi_{in,\beta}^2 + 2m\pi. \quad (23)$$

To have the particle beam cooling occurred in the first transverse radial region of phase space, corresponding to the condition $J_0(\Delta \varphi_{in,\beta,m}^1) > 0$ the amplitude of betatron oscillations of the beam according to (20) must satisfy the condition $\Delta \varphi_{in,\beta,m}^1 < \mu_2 = 2.4$, i.e. not exceed the value (see (12) for definition of $B_x(0)$)

$$A_x(0) < 2.4 / k_{1c} B_x(0) \sim 0.4 \lambda_c / B_x(0). \quad (24)$$

From (24) it follows that to work with particle beams having amplitudes of betatron oscillations $A_x(0) \sim 1 \text{mm}$ at a wavelength of undulator radiation $\lambda_c \sim 10^{-4} cm$ one should make a value of $B_x(0)$ as small as $\sim 10^{-3} cm$, i.e. make the magnetic structure of the cooling storage ring close to no dispersive ($B_x(0) \sim c_x \sim d_x << 1$, see (12)).

### b. Beam cooling in a transverse phase space.

Betatron oscillations of particles lead to the length increase of its path between the signal and kicker undulators and to corresponding phase shift of particle's entry into the UWP at the entrance into the kicker undulator $\Delta \varphi_{in,\beta}(s_{kik}, n)$ relative to the phase of particles having the same energy but with zero amplitude of betatron oscillations. This shift for a given amplitude of betatron oscillations of the particle in the general case depends on the initial deviation from the instantaneous particle orbit at the entrance to the signal undulator $x_\beta(s = 0, n)$, $z_\beta(s = 0, n)$, (on the turn number $n$, see (5), (12)). Hence, according to (14), it follows that the jump in energy $\Delta \varepsilon_{jp}(\Delta \varphi_{in})$ of particles, and hence the position of the instantaneous orbit jump $\Delta x_\eta(s_{kik}, n) = \eta_x(s_{kik}, n) \beta^{-2} \Delta \varepsilon_{jp}(\Delta \varphi_{in}) / \varepsilon$ equal to the

$$\Delta x_\eta(s_{kik}, n) = -\eta_x(s_{kik}, n) \beta^{-2} \varepsilon^{-1} P_{loss}^{max} T f(\varphi_{in}) \cos [\Delta \varphi_{in,\eta} + \Delta \varphi_{in,\beta}^1(n) + \Delta \varphi_{in,\beta}^2] \quad (25)$$

through the entry phase shift $\Delta \varphi_{in,\beta}^1(n)$ depend on the initial deviation of the particle at the entrance to the undulator signal. Roughly, on the same amount the amplitude of the betatron oscillations of the particle is changed [4], [8].

For cooling of particle beam in the transverse phase space it is necessary to create conditions under which the jumps of instant particle orbits decelerated at the entrance to the kicker undulator for negative deviations from their instant orbits were on average more jumps of instant particle orbits are at positive deviations from these orbits. Then the particle, having a negative deviation from the instantaneous orbit, at the first jump of the instantaneous orbit position, will be closer to it than to be retired from it, having a positive deviation. In this case, each pair of such jumps will lead to a decrease in the amplitude of radial betatron oscillations and the beam will be cooled as a whole.

The difference between amplitudes of betatron oscillations of irregular particles corresponding positive and negative deviations from its instantaneous orbit $x_\beta(s = 0, k) = -x_\beta(s = 0, l) \geq 0$, i.e. in the case where it is equivalent to the same values and the difference marks entry phase particles in the UWP at the entrance to the kicker undulator $\Delta\varphi^1_{in,\beta}(k) = -\Delta\varphi^1_{in,\beta}(l) \geq 0$, determine the magnitude of change in the oscillation amplitude for a couple of jumps of the particle energy

$$\Delta A(s_{kik}) \sim \Delta x_\eta(s_{kik}) = -\eta_x(s_{kik})\beta^{-2}\varepsilon^{-1}P_{loss}^{max}Tf(\varphi_{in})$$
$$[\cos(\Delta\varphi_{in,\eta} + \Delta\varphi^1_{in,\beta}(k) + \Delta\varphi^2_{in,\beta}) - \cos(\Delta\varphi_{in,\eta} + \Delta\varphi^1_{in,\beta}(l) + \Delta\varphi^2_{in,\beta})] =$$
$$\eta_x(s_{kik})\beta^{-2}\varepsilon^{-1}P_{loss}^{max} T f(\varphi_{in})\sin(\Delta\varphi_{in,\eta} + \Delta\varphi^2_{in,\beta})\sin(\Delta\varphi^1_{in,\beta}(k)),$$

where now the values left with the phases of the quantities $\Delta\varphi^1_{in,\beta}(k) = \Delta\varphi^1_{in,\beta,m}\cos[\alpha_x + b_x + 2\pi k\nu_x] \geq 0$. Averaging over a large number of pairs of instantaneous orbit jumps $K = N/2 \gg 1$ expression $\sin(\Delta\varphi^1_{in,\beta}(k))$ which is equal $<\sin(\Delta\varphi^1_{in,\beta}(k))> = (1/K) \sum_1^K <\sin[\Delta\varphi^1_{in,\beta,m}\cos(\alpha_x + b_x + 2\pi k\nu_x)]>$, as in the previous case can be simplified by leaving it in the argument of the phases $\alpha_x + b_x + 2\pi k\nu_x$ terms $\xi \leq 2\pi$, which will include a range of phase $-\pi/2 \leq \xi \leq \pi/2$ intervals $\pi/K$. In this case, the sum can be replaced by integral $(1/\pi)\int_{-\pi/2}^{\pi/2}\sin[\Delta\varphi^1_{in,\beta,m}\cos\xi]d\xi = \pi H_0(\Delta\varphi^1_{in,\beta,m})$, where $H_0(\Delta\varphi^1_{in,\beta,m})$ - is a Struve function of zeroth order ($H_0(0) = H_0(4.34) = 0$, $H_0(1.98) \sim 0.79$). Hence the expression for the average rate of change in the amplitude of the betatron oscillations

$$dA/dt = \Delta A(s_{kik})/T = \eta_x(s_{kik})\beta^{-2}\varepsilon^{-1}P_{loss}^{max} f(\varphi_{in})\sin(\Delta\varphi_{in,\eta} + \Delta\varphi^2_{in,\beta})H_0(\Delta\varphi^1_{in,\beta,m}). \quad (26)$$

From (26) it follows that for $H_0(\Delta\varphi^1_{in,\beta,m}) > 0$, i.e. at amplitudes entry phase particles in the UWP at the entrance into the kicker undulator satisfying $0 < \varphi^1_{in,\beta,m} < \mu_1$ $\mu_{2+2i} < \varphi^1_{in,\beta,m} < \mu_{3+2i}$, $i = 1, 2, 3...$ in a phase diapason $-\pi + 2m\pi < \varphi_{in,\eta} + \Delta\varphi^2_{in,\beta} < 0 + 2m\pi$ the particle beam cooling is going in the transverse phase space (by the radial angle). Here - $\mu_j$ is the root of Struve function. This means that the conditions for particle beam cooling simultaneously in the longitudinal and first transverse radial parts of the phase space region coincide only in limited areas $m$ of the longitudinal range of phases $-\pi + 2m\pi < \varphi_{in,\eta} < 0 + 2m\pi$.

The rate of change of the amplitude of the betatron oscillations (26) for small amplitudes of betatron phase $\Delta\varphi^1_{in,\beta,m} \sim A_x(s)B_x(0)$) and small deviations $\varphi_{in,\eta,m} = -\pi/2 + 2m\pi$ from the equilibrium phase is proportional to the $\eta_x(s_{kik})\Delta\varphi^1_{in,\beta,m} = \eta_x(s_{kik})k_{1c}A_x(0)B_x(0)$. The average rate of change of the particle's energy in the storage ring for small deviations from the equilibrium energy of the particle is proportional to the value

$$\cos(\varphi_{in,\eta} + \Delta\varphi^2_{in,\beta})\Big|_{\varphi_{in,\eta} \to \pi/2} J_0(\Delta\varphi_{in,\beta,m})\Big|_{\Delta\varphi_{in,\beta,m} \ll 1} \sim (\pi/2 - \varphi_{in,\eta}) \ll 1$$

For optimal performance of the method TTOSC, the phase space region occupied by the particle beam, should be positioned in the first energy region and the first phase zone adjacent to the axis $s$ (radial and vertical). By selecting the appropriate value of the local slip factor, the coefficient $B_x(0)$, the magnitudes and signs of the dispersion function in sections of pickup and kicker undulator, and by coupling of vertical and horizontal betatron oscillations it becomes possible to cool the particle beam in the vertical plane. In this case, the particle beam with a spread of betatron oscillation amplitudes defined by expression $\Delta\varphi^1_{in,\beta,m} = k_{1c}A_x(0)B_x(0) < \mu_1$ and energy spread of the relevant range of angles $-\pi < \Delta\varphi_{in,\eta} < 0$, according to (20), (26) will be concentrated in the region of equilibrium phase $\varphi_{in,\eta,1} = -\pi/2$ and the corresponding energy $\varepsilon_{r,1}$ (see Fig. 2). In this case, closer the initial phase $\varphi_{in,\eta}$ to the equilibrium phase particles and smaller the amplitude of the betatron oscillations of the particles, – the lower is the rate of convergence of the particle energy with an equilibrium energy and lower the rate of decrease of betatron oscillation is (see (20), (26)).

The transport system of the undulator beam from the pickup to kicker undulator configured so that at the beginning of the cooling process, the energy spread of the beam corresponds to a range of phases $-\pi + 2m\pi < \varphi_{in,\eta} + \Delta\varphi_{in,\beta}^2 < 0 + 2m\pi$ at amplitudes of entry phase particles in the UWP at the entrance to the kicker undulator satisfying the condition $0 < \varphi_{in,\beta,m}^1 < \mu_1$, $\mu_{2+2i} < \varphi_{in,\beta,m}^1 < \mu_{3+2i}$, $i = 1,2,3$. UWP amplitude selected maximum amplitude for the selected OA, does not depend on the current situation, so as to maximize the rate of cooling. It is assumed that the cooling time is much greater than the orbital period of particles in the drive. The cooling rate of the particle beam is inversely proportional to the amplitude of the UWP. Therefore the pickup undulator field should be chosen closer to the optimal one (undulator strength parameter).

The amplitude of UWP selected as a maximum one for the chosen OA, which does not depend on the current moment, so as to maximize the rate of cooling. It is assumed that the cooling time is much greater than the orbital period of particles in the ring. The cooling rate of particle beam is inversely proportional to the amplitude of the UWP. Therefore the pickup undulator field should be chosen closer to the optimal (undulator strength parameter $K \sim 1$).

**B. Cooling with RF turned on.**

In the method of TTOSC, the equilibrium particle with respect to its amplified UWP should be both relatively synchronous with the driving RF voltage at the resonator. In this case, the particle beam is cooled to about as well as for the RF voltage is switched off. In the initial state, the entire beam must be in the first half of the energy band in the longitudinal phase space between the phases $-\pi < \varphi_{in,\eta} < 0$.

Energy and amplitude of the betatron oscillations loss rate of the beam particles in this method of cooling (at a constant gain OA) will drop with decreasing of angular and energy spread of the beam (amplitudes of the betatron and phase oscillations). This is the main disadvantage of TTOSC in comparison with EOC.

## 5. METHOD OF ENHANCED OPTICAL COOLING (EOC)

We noted above that in the method TTOSC, the interaction of particle in a kicker undulator with the half of UWP emitted by particles in the pickup undulator and amplified in the OA, increases the amplitude of the betatron oscillations of the particles. The amplitude attenuation effect of betatron oscillations of the particles in this case arises in some regions of phase space due to the difference of average values of positive and negative jumps of instant particle orbits in these areas.

In the cooling method EOC the path for UWP which lead to increase in the amplitude of the betatron oscillations of particles, is overlapped by the movable screen mounted in the image plane of the optical system drive. On this plane the distribution of particles in the pickup undulator is projected, which contains the information about direction of the deviation of these particles on their instant orbits radially and vertically and allowing selection emitted UWP by this screen. The distance between the signal and kicker undulators is chosen so that the phase shift of the betatron oscillations of particles equal value $\varphi_{pk}^b = 2\pi(k_{pk} + \frac{1}{2})$ where $k_{pk} = 0,1,2,3...$ - are integers. In this case, the particle having in a pickup undulator positive deviation from its instantaneous orbit in the kicker undulator will have negative deviation - a necessary condition for the cooling of particle beam.

### A. EOC with RF turned off.

In selecting of UWP in EOC the value

$$\langle \cos\varphi_{in} \rangle \sim (1/2N)\sum_{n=1}^{N}\cos[\varphi_{in,\eta} + |\Delta\varphi^1_{in,\beta}| + \Delta\varphi^2_{in,\beta}] =$$

$$(1/2N)\sum_{n=1}^{N}\{\cos[\varphi_{in,\eta} + \Delta\varphi^2_{in,\beta}]\cos|\Delta\varphi^1_{in,\beta}(n)| - \sin[\varphi_{in,\eta} + \Delta\varphi^2_{in,\beta}]\sin|\Delta\varphi^1_{in,\beta}(n)| =$$

$$(1/2N)\cos[\varphi_{in,\eta} + \Delta\varphi^2_{in,\beta}]\sum_{n=1}^{N}\cos|\Delta\varphi^1_{in,\beta}(n)| - (1/2N)\sin[\varphi_{in,\eta} + \Delta\varphi^2_{in,\beta}]\sum_{n=1}^{N}\sin|\Delta\varphi^1_{in,\beta}(n)|$$

$$= 0.5\{\cos[\varphi_{in,\eta} + \Delta\varphi^2_{in,\beta}]J_0(\Delta\varphi^1_{in,\beta,m}) - \sin[\varphi_{in,\eta} + \Delta\varphi^2_{in,\beta}]H_0(\Delta\varphi^1_{in,\beta,m})\}.$$ Factor $(1/2N)$ takes into account the overlap screen half UWP emitted particles in the pickup undulator.

Thus, the average rate of change of the particle energy (8) and the mean value of jumps of particle energy in the storage for arbitrary values of the betatron oscillation amplitudes take the form

$$\overline{P}_{loss} = -0.5 P_{loss}^{max} f(\varphi_{in})\{\cos(\varphi_{in,\eta} + \Delta\varphi^2_{in,\beta})J_0(\Delta\varphi^1_{in,\beta,m}) - \sin[\varphi_{in,\eta} + \Delta\varphi^2_{in,\beta}]H_0(\Delta\varphi^1_{in,\beta,m})\},$$

$$\overline{\Delta\varepsilon}_{jp} = -P_{loss} T. \qquad (27)$$

Accordingly, the rate of change of the amplitude of the betatron oscillations of the particle, according to (25) and (27) will be

$$dA/dt = -0.5\eta_x(s_{kik})\beta^{-2}\varepsilon^{-1}P_{loss}^{max} f(\varphi_{in})\{\cos(\Delta\varphi_{in,\eta} + \Delta\varphi^2_{in,\beta})J_0(\Delta\varphi^1_{in,\beta,m}) -$$

$$\sin(\varphi_{in,\eta} + \Delta\varphi^2_{in,\beta})H_0(\Delta\varphi^1_{in,\beta,m})\},. \qquad (28)$$

Various embodiments of the EOC in this mode are possible.

**a)** The pickup undulator installed in the straight section of storage ring having zero dispersion function. Instantaneous orbits of the beam particles in this interval coincide, and beam sizes are determined by the amplitudes of betatron oscillations and do not depend on the energy spread of the beam. The screen stands still and opens the image of particles created in the image plane by theirs UWP emitted in the pickup undulator and having the only positive deviations in radius. Cooling particles in this embodiment EOC will occur in all areas of energy and transverse radial zones of the phase space. This option is a preferred one for rapid cooling of the betatron amplitudes in the particle beam.

**b)** The pick-up undulator installed in the straight section having nonzero dispersion function. In this case, the image size of the beam is defined as the amplitudes of betatron oscillations of particles and energy spread of the beam. The screen moves so that it opens first the way to UWP emitted by particles having a large energy and positive deviations from their instant orbits. In the kicker undulator, installed in the straight section of storage ring, the deviation of particles from the instantaneous orbit at the same turn is negative. Therefore, the energy loss by the beam particles in this case is always accompanied by convergence of energy with energy corresponding to the equilibrium energy state $\varepsilon_m$, phases - with stable equilibrium phases $\varphi_m = 0$ and simultaneously decreasing amplitude of betatron oscillations. Screen stops at a position corresponding to the position of instantaneous orbit having a desired energy $\varepsilon_g$ stays there for a while, and then returns to the initial position. This process is repeated periodically. Desired energy corresponds to the phase $\varphi_{in,\eta,g} \sim -\pi/4$ in which the rate of energy loss and reduction of betatron amplitudes, and hence the cooling rate of the particle beam can be orders of magnitude greater than the rate of cooling beams in phases $\varphi_g \sim \varphi_m \sim -\pi/2$ which characterizes the method TTOSC.

### B. EOC with RF turned on.

Various embodiments of the EOC in this mode are possible.

**a)** Pickup undulator installed in the straight section of damping ring with big dispersion and with small beta function. Equilibrium with respect to UWP energy of the particle $\varepsilon_r$ corresponding to the equilibrium phase $\varphi_{in,\eta,r} = -\pi/2$ is set below the synchronous particle energy $\varepsilon_s$ to the voltage on the RF cavity of the ring. The screen covers UWP emitted by the particles moving in instant orbits corresponds to the energy of these particles $\varepsilon < \varepsilon_s$. The kicker undulator installed in the straight section of the storage ring which also has a big dispersion and a small beta function at this location.

In this case, the particles moving at the top of the longitudinal phase space bounded by a separatrix ($\varepsilon > \varepsilon_s$) become decelerated in the electric field of UWP, and their amplitude of phase and betatron oscillations are reduced. In the lower region ($\varepsilon < \varepsilon_s$), the interaction of particles with UWP and thus the damping or buildup of the phase and betatron oscillations is missing. Therefore, the energy loss rate and the amplitude of betatron oscillations of the beam particles remain to be close to the maximum.

**b)** Pick-up undulator installed in the straight section of damping ring with zero dispersion and large beta function. Equilibrium with respect to the UWP energy of the particle $\varepsilon_r$ corresponding to the equilibrium phase $\varphi_{in,\eta,r} = -\pi/2$ is set below the synchronous particle energy $\varepsilon_s$ to the voltage on the RF cavity of the ring. The screen covers UWP emitted by the particles moving in instant orbits corresponding to the energy $\varepsilon < \varepsilon_s$ of these particles. The kicker undulator installed in a straight section of the ring having large dispersion and small beta function.

In this case, the particles moving at the top of the longitudinal phase space bounded by a separatrix ($\varepsilon > \varepsilon_s$) decelerated in the electric field of UWP, and theirs amplitudes of betatron and phase oscillations are reduced. At the bottom of separatrix ($\varepsilon < \varepsilon_s$) interaction of particles with UWP and, hence, an attenuation or the buildup phase and betatron oscillations are absent. Therefore, the energy and amplitude of the particle betatron oscillations loss rate in the beam remain close to the maximum possible.

EOC disadvantage is limitation on the accuracy of measuring the position of the particles in the pickup undulator associated with the diffraction limit in electrodynamics [4].

## 6. OSC WITH QUADRUPOLE PICKUP UNDULATOR

### a. Radiation of a particle in a quadrupole undulator.

In the OSC scheme [2], a quadrupole undulator used as a pick-up undulator. In this undulator the transverse component of magnetic field near the axis is a linear function of the transverse coordinate and alternating function of the longitudinal coordinate *s*. The conventional system of quadrupole lenses with alternating polarity could serve as a quadrupole undulator. In this case, the focal length of each lens must be bigger than the distance *L* between them. Selection of the desired optical signal component in this case could be accomplished by using the polarizer.

Thus, bit shifted transversely by the amount (1)

$$x(s) = x_\beta(s) + x_\eta(s)$$

the particle occurs in the undulator field of the form $H_y(x) = g \cdot x$ or $H_x(y) = g \cdot y$ where $g$ – is gradient magnetic field lens. For a particle moving with an offset *x* relative to the central path, the field looks like an ordinary dipole undulator field with an amplitude proportional to the displacement of the particle. Therefore, the power of radiation of the beam in a quadrupole wiggler at any current moment of cooling is less than at the beginning of the cooling process.

The number of photons of undulator radiation emitted by particle in the spectral band $\Delta\omega/\omega_{1c} = 1/M$

$$N_\gamma \cong 4\alpha K^2/(1+K^2) = 4\alpha(A/A_0)^2/[1+(A/A_0)^2], \qquad (29)$$

where $K^2 \sim x^2$ and $A_0$ corresponds to the initial amplitude, $A$ is current amplitude. Each of the emitted photons carries energy

$$\hbar\omega \cong \frac{2\pi\hbar c\gamma^2}{L\cdot(1+K^2)}, \qquad (30)$$

and total energy emitted by the particles in the UWP is $W_{pick} \cong \hbar\omega N_\gamma$.

Feature of a quadrupole wiggler (undulator) is that the center frequency of the radiation (30) is a function of the betatron amplitude. Expression (30) shows that if the initial amplitude matches $K \cong 1$, the bandwidth of optical amplifier should overlap the octave at the end of the cooling cycle when $K \rightarrow 0$.

—**b. Beam cooling in a transverse phase space.**

By idea of stochastic cooling, the electromagnetic wave emitted by a particle-UWP should interact with the *same* particle. This can be done if the UWP and the beam are moving in the same direction. Decrease in the amplitude of the betatron oscillations, as in the case usual undulator through a change of the particle's energy in the place where the dispersion of the trajectory has a nonzero value $\eta(s) > 0$, see Figure 2. This last means that the amplitude of the transverse betatron oscillations on the new closed orbit[3] changes by an amount $\Delta x \cong \eta(s)\Delta\varepsilon/\varepsilon_s$, where $\Delta\varepsilon/\varepsilon_s$ is the relative change in energy, see Figure 2.

## 7. CONCLUSIONS

In this paper we presented a theory of optical stochastic cooling of particle beams in storage rings in the low current approximation. It was shown that during this cooling procedure the six-dimensional space occupied by the particle beam is generally broken into regions (zones) with theirs own centers of equilibrium states [4]. In each area there is going a beam cooling. As a result, the beam breaks the coordinates, energy and angles into different cooled regions separated by free phase space. It is shown that for deep cooling of beam it is necessary to create conditions for cooling in three directions. In this case the radial and vertical betatron oscillations could be coupled, or to modify the storage ring for including the orbital slots which extending the trajectory in the vertical plane.

In some cases it is desirable to create conditions for the simultaneous cooling of the entire particle beam. To do this, one can pre-arrange it in one region of phase space by adjusting accordingly the magnetic structure of the cooling system of the storage ring.

Cooling methods allow one to split a chilled beam on $\sim 2M$ equidistant energy layers that are further in the undulator and in a field of strong electromagnetic wave can be converted into a modulated beam density on $\sim 2M$ -the harmonica of laser wave [15] - [16]. Such beams can be used effectively in the DUV and X-ray free-electron lasers [15] - [18].

The main disadvantage of TTOSC method is in reduction of the rate of energy and amplitude of betatron oscillations of particles change with their approach to equilibrium. They determine the decay time of the angular and energy spread of the beam, its equilibrium size determined by the destructive influence of the quantum fluctuations of synchrotron radiation and scattering inside the beam. EOC method has the rate of cooling higher than TTOSC.

---

[3] Which corresponds to the new equilibrium orbit.